\documentclass[%
 aip,
 amsmath,amssymb,
 reprint,%
]{revtex4-1}

\usepackage{graphicx}
\usepackage{dcolumn}
\usepackage{bm}
\usepackage{bbold}
\usepackage{amsmath}
\usepackage{comment}
\usepackage{xcolor}

\begin{document}

\preprint{APS/123-QED}

\title{Experimental evidence of enhanced broadband transmission in disordered systems with mirror symmetry}

\author{Matthieu Davy}
\affiliation{%
Univ.\ Rennes, CNRS, IETR (Institut d'{\'E}lectronique et des Technologies du num{\'e}Rique), UMR–6164, F-35000 Rennes, France
}%

\author{Cl{\'e}ment Ferise}
\affiliation{%
Univ.\ Rennes, CNRS, IETR (Institut d'{\'E}lectronique et des Technologies du num{\'e}Rique), UMR–6164, F-35000 Rennes, France
}%

\author{{\'E}lie Ch{\'e}ron}
\affiliation{
Laboratoire d'Acoustique de l'Universit{\'e} du Mans (LAUM), UMR 6613, Institut d'Acoustique - Graduate School (IA-GS), CNRS, Le Mans Universit{\'e}, France
}%

\author{Simon F{\'e}lix}
\affiliation{
Laboratoire d'Acoustique de l'Universit{\'e} du Mans (LAUM), UMR 6613, Institut d'Acoustique - Graduate School (IA-GS), CNRS, Le Mans Universit{\'e}, France
}%

\author{Vincent Pagneux}
\affiliation{
Laboratoire d'Acoustique de l'Universit{\'e} du Mans (LAUM), UMR 6613, Institut d'Acoustique - Graduate School (IA-GS), CNRS, Le Mans Universit{\'e}, France
}%

%


\begin{abstract}
We demonstrate in microwave measurements the broadband enhancement of transmission through an opaque barrier due to mirror symmetry. This enhancement relies on constructive interference between mirror scattering paths resulting from strong internal reflections at the left and right interfaces of a multichannel cavity. We observe a strong sensitivity of the conductance to a shift of the barrier from the center of the cavity. Remarkably, the impact of mirror symmetry can be further increased by tuning the degree of disorder within the cavity. We report an additional enhancement of the conductance found by symmetrically placing randomly located scatterers. Our results illuminate the impact of symmetry and disorder correlation on transmission through complex systems.

\end{abstract}

\maketitle

Understanding interference phenomena is essential to characterize the transmission of waves through scattering systems. In diffusive samples, the interference of scattering paths is \textit{a priori} random so that the average transmitted intensity and the profile of the energy density can be predicted by the diffusion equation. Deviation from the diffusion theory however arise in periodic and disordered scattering systems as a result of constructive or destructive interference between scattering paths. By manipulating the incident wavefront, the transmission may be fully controlled as the distribution of transmission eigenvalues spans from zero (closed channels) to unity (open channels) in diffusive samples \cite{Dorokhov1984,Imry1986,Mello1988,Beenakker1997,Vellekoop2008a,Kim2012,Goetschy2013,Gerardin2014,Sarma2016,Rotter2017,Sarma2017,Yilmaz2019}. The spatial correlation and the strength of the disorder within the sample can also be specifically engineered to give rise to fascinating interference effects. A well-known example is the formation of band gaps in photonic crystals  \cite{Yablonovitch1993}. In strongly disordered samples, the average transmission is coherently suppressed in the regime of Anderson localization \cite{Akkermans2007,Lagendijk2009}. Transmission may also be substantially enhanced by tuning the degree of correlation of the disorder. Stealth hyperuniform media below a threshold frequency are transparent to incoming radiations at densities for which an uncorrelated disorder would be opaque \cite{Leseur2016,Aubry2020}. 

Robust interference phenomena can also be induced by a mirror symmetry within a cavity or a disordered medium. Whitney \textit{et al.} demonstrated that the conductance through an opaque barrier placed within a symmetric quantum dot is greatly enhanced as a result of constructive interference between symmetric classical paths \cite{Whitney2009,Whitney2009a}. This broadband effect is reminiscent of coherent backscattering for reflected waves \cite{Albada1985,Akkermans2007} and coherent forward scattering in localized samples \cite{Karpiuk2012} that are robust to a statistical averaging. A significant broadband enhancement has also been reported in diffusive waveguides with open boundary conditions at the left and right interfaces \cite{Cheron2019}. Transmission through a random but symmetric diffusive slab with an opaque barrier in the middle can indeed be much larger than transmission through the barrier alone, with a strong modification of the distribution of transmission eigenvalues. Instead of being limited to a maximal value imposed by the barrier strength, this distribution in symmetric disorders coincide with its expectation for random configurations in absence of the barrier. Open channels with transmission eigenvalues close to unity are especially recovered. In addition, a deep subwavelength sensitivity of the conductance to a shift of the barrier or to symmetry defects in its surrounding disorder has been reported \cite{Whitney2009,Cheron2019,Cheron2020,Cheron2020_scirep}. Nevertheless, theoretical studies have been confirmed only by numerical simulations and a clear experimental demonstration of the impact of left-right symmetry is still missing. 

\begin{figure}
\includegraphics[width=8.5cm]{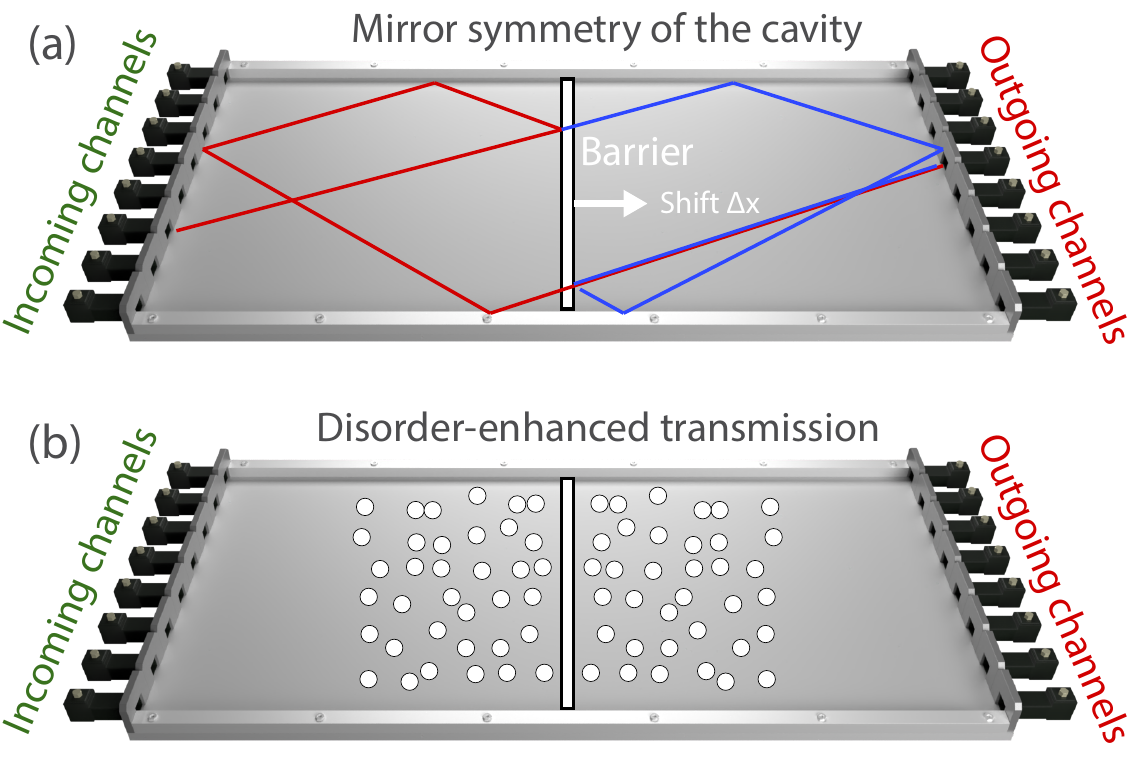}
\caption{\label{fig:fig1} Sketch of the cavity for which a barrier (an aluminum bar) is placed in the middle to create a left-right mirror symmetry. The top plate has been removed to see the interior of the cavity. Measurements of the transmission matrix $t(\nu)$ between two arrays of $N=8$ single-channel waveguides are carried out in the microwave range using two electromechanical switches connected to two ports of a Vector Network Analyzer. (a) Red and dashed blue lines illustrate the superposition of two paths with a mirror symmetry that coherently enhance the transmission. To explore the impact of left-right symmetry, the barrier is shifted from the center of the cavity by $\Delta{x}$. (b) In the same device with a centered barrier, a random but symmetric distribution of aluminium cylindrical scatterers is placed on both sides of the barrier.}
\end{figure}

In this article, we evidence experimentally the broadband enhancement of transmission due to the mirror symmetry in a multichannel cavity in which a barrier is placed. First, we investigate the sensitivity of the conductance to the mirror symmetry for an empty cavity by progressively shifting the barrier from the center. We clearly observe that the conductance is maximum in the symmetric configuration. The coherent interference of scattering paths is illustrated by the temporal variations of the transmitted intensity.  Second, we add a symmetric disorder within the cavity and report a maximal enhancement by a factor three of the conductance.\\

Our experimental setup is a multichannel cavity of length $L = 0.5$~m, width $W = 0.25$~m and height $h=8$~mm (see Fig.~\ref{fig:fig1}(a)). The cavity is effectively two-dimensional as a single vertically polarized mode can propagate. Spectra of the complete transmission matrix (TM) $t(\nu)$ are measured between two arrays of $N = 8$ antennas. These antennas are single waveguide channels fully coupled to the system between 11 and 17~GHz that are attached to the cavity at the left and right interfaces (see Fig.\ref{fig:fig1}) \cite{Davy2021}. The TM is built upon the field transmission coefficients $t_{ba}(\nu)$ between each incoming antenna $a$ and outgoing antenna $b$. We stress that strong internal reflections at the interfaces of the cavity result from metallic boundary conditions at the spacing between the antennas. All openings of the cavity are controlled with transmitting or receiving antennas and the TM is therefore complete. 

\begin{figure}
\includegraphics[width=8.5cm]{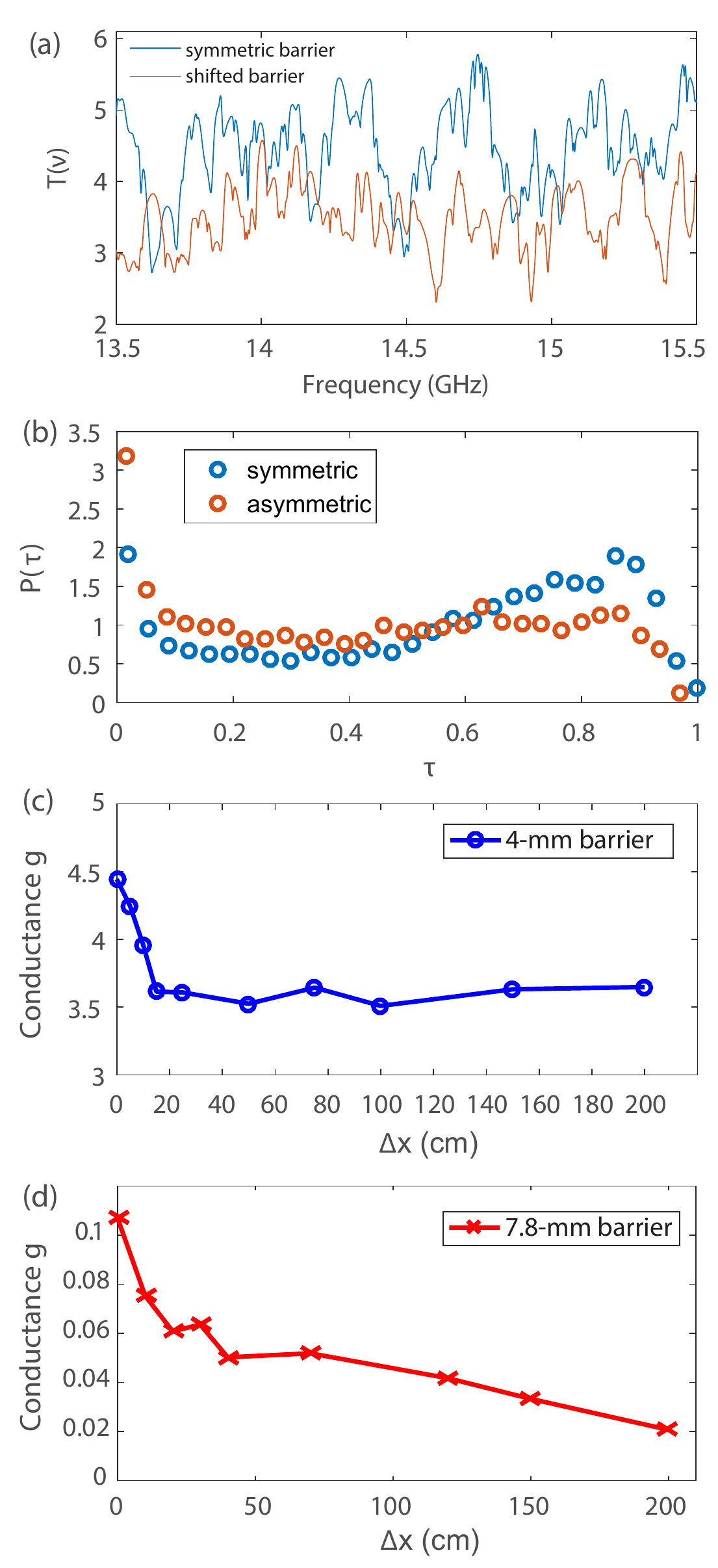}
\caption{\label{fig:fig_conductance} (a) Spectrum of the transmittance $T(\nu)$  for a 4~mm high barrier placed symmetrically within the empty cavity (blue line) and then shifted by $\Delta{x} = 50$~mm (red line). (b) Corresponding distributions of transmission eigenvalues $P(\tau)$ for a mirror symmetry (blue dots) and broken mirror symmetry (red dots). (c) Variations of the conductance with the shift $\Delta{x}$ of the barrier. (d) Same as (c) but for a barrier with stronger reflectivity of height equal to 7.8~mm.}
\end{figure}

We first explore the impact of left-right mirror symmetry on the conductance with a single barrier which is a 4~mm height metallic rectangular bar. The barrier is first placed symmetrically to the left and right interfaces ($\Delta{x} = 0$) and then shifted from the center by $\Delta{x} = 50$~mm. The spectrum of the transmittance, which is the sum of the total transmsission over all the incident channels $T(\nu) = \Sigma_{ba} |t_{ba}(\nu)|^2 $ is seen to be larger in the symmetric configuration. The dimensionless conductance $g = \langle T(\nu) \rangle$ found from an averaging over the frequency range is given by $g= 4.44$ for $\Delta{x} = 0$ and  $g  = 3.52$ for $\Delta{x} = 50$~mm (see Fig.~\ref{fig:fig_conductance}(a)). 

The transmittance may also be expressed in terms of the $N$ transmission eigenvalues $\tau_n(\nu)$ of $t^\dagger(\nu) t(\nu)$,  $T(\nu) = \Sigma_{n=1}^N \tau_n(\nu) $ \cite{Imry1999}. The distribution $P(\tau)$ is found to be bimodal, as expected for multichannel cavities \cite{Baranger1994,Jalabert1994,Rotter2017}, with two peaks corresponding to closed channels ($\tau \sim 0$) and open channels ($\tau \rightarrow 1$) (see Fig.~\ref{fig:fig_conductance}(b)). This distribution highlights that such a sample can be either opaque or almost transparent to incoming radiations depending on the incident wavefront. Note that the peak for open channels is here found at $\tau = 0.9$ instead of $\tau = 1$ as a result of small dissipation within the sample \cite{Goetschy2013,Davy_in_prep}. As illustrated with numerical simulations of random asymmetrical media in Supplementary Material, small losses within the sample indeed leads to a shift of the second characteristic peak towards smaller transmission but does not suppress it \cite{Yamilov2016}. As the configuration becomes asymmetric ($\Delta{x} \neq 0$), the amplitude of this peak is further reduced leading to smaller values of $g$.

The conductance shown in Fig.~\ref{fig:fig_conductance}(c) decreases rapidly with increasing $\Delta{x}$ between $\Delta{x} = 0$ and $\Delta{x} = 15$~mm, and for $\Delta{x} > 15$~mm, it reaches a plateau. This shift is of the order of $\lambda/1.5 = 10.7$~mm at a frequency of $f=14$~GHz. We then increase the barrier's reflectivity by using a metallic bar with height of $7.8$~mm. The barrier now almost fully fills the height of the system. The reduction of $g$ with $\Delta{x}$ is even more significant  as it almost reaches an order of magnitude, from $g=0.11$ for $\Delta{x} = 0$ to $g=0.002$ for $\Delta{x} = 200$~mm (see Fig.~\ref{fig:fig_conductance}(d)).

The origin of the enhancement of $g$ in a symmetrical configuration can be understood from the schematic view of two scattering paths for a symmetric barrier given in Fig.~\ref{fig:fig1}, as shown in Ref.~\cite{Whitney2009}. For an incident wave impinging on the barrier with an angle $\theta$, the transmitted field to an outgoing channel can be split into two mirror scattering paths. The first path $p_1$ is successively reflected by the barrier (coefficient $r_\textup{B}(\theta)$), reflected at the left interface in the spacing between two antennas and finally transmitted through the barrier (coefficient $t_\textup{B}(\theta')$) before being absorbed at the right receiving channel (see red path on Fig.~\ref{fig:fig1}(a)). Its contribution for a path length $L_1$ is $\psi_1 = r_\textup{B}(\theta) t_\textup{B}(\theta') e^{\textup{i}kL_1}$ ($k$ is the wave number). The second scattering path $p_2$ of length $L_2$ is first transmitted through the barrier and then follows a path which is the mirror of $p_1$ at the right side of the cavity, $\psi_2 = r_\textup{B}(\theta') t_\textup{B}(\theta) e^{\textup{i}kL_2}$. For a mirror symmetry giving $L_1 = L_2$, the intensity $\langle I \rangle = \langle |\psi_1 + \psi_2|^2 \rangle= \langle | r_\textup{B}(\theta) t_\textup{B}(\theta') + r_\textup{B}(\theta') t_\textup{B}(\theta) |^2 \rangle$ is significantly enhanced relative to its average $\langle I \rangle = \langle | r_\textup{B}(\theta) t_\textup{B}(\theta')|^2\rangle + \langle |r_\textup{B}(\theta') t_\textup{B}(\theta) |^2 \rangle$ found when $L_1$ and $L_2$ are independent random variables.

\begin{figure}
\includegraphics[width=8.5cm]{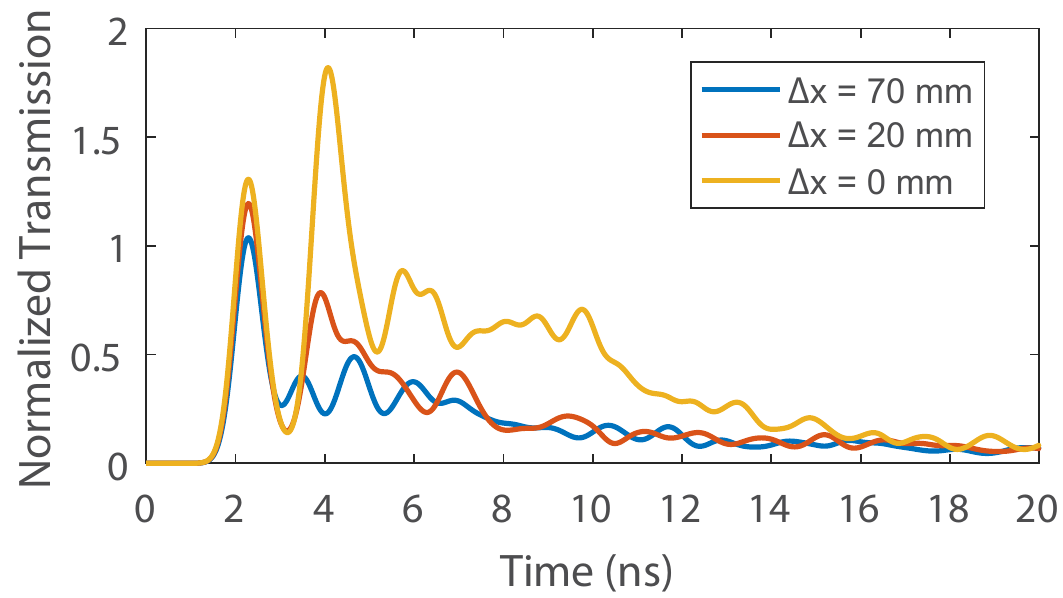}
\caption{\label{fig:fig_time}. Variations of the transmission in the time domain for a shift $\Delta x=7$ cm (blue line), $\Delta x=2$ cm (orange line) and $\Delta x=0$ cm (yellow line). The amplitude of the second pulse at $t = 4.15$~ns is clearly enhanced by a mirror symmetry.}
\end{figure}

To illustrate the impact of constructive interference, we consider the time variation of transmitted field $\Tilde{T}(t) = \Sigma_{ba} |\Tilde{t}_{ba}(t)|^2 $ in Fig.~(\ref{fig:fig_time}). The field transmission coefficients in the time domain $\Tilde{t}_{ba}(t)$ are obtained from the inverse Fourier transform of the elements $t_{ba}(\nu)$ for a Gaussian pulse of central frequency $f_0=14.5$ GHz and bandwidth $\Delta f=400$ MHz. We compare $\Tilde{T}(t)$ for $\Delta x=0$, $\Delta x=20$~mm and $\Delta x=70$~mm. The magnitude of the first pulse found at $t_0=2.1$~ns weakly depends on $\Delta x$ as the ballistic wave through the cavity is barely impacted by the position of the barrier. However, the second pulse associated to the double scattering illustrated in Fig.~\ref{fig:fig1}(a) is nicely enhanced by the mirror-symmetry with a magnitude which even exceed the one of the direct pulse for $\Delta x=0$. At late times, the pulses associated to multiple scattering between the barrier and the interfaces are mixed and therefore cannot be resolved temporally but $\Tilde{T}(t)$ for $\Delta x =0$ dominates the other curves until $t = 20$~ns.\\

\begin{figure}
\includegraphics[width=8.5cm]{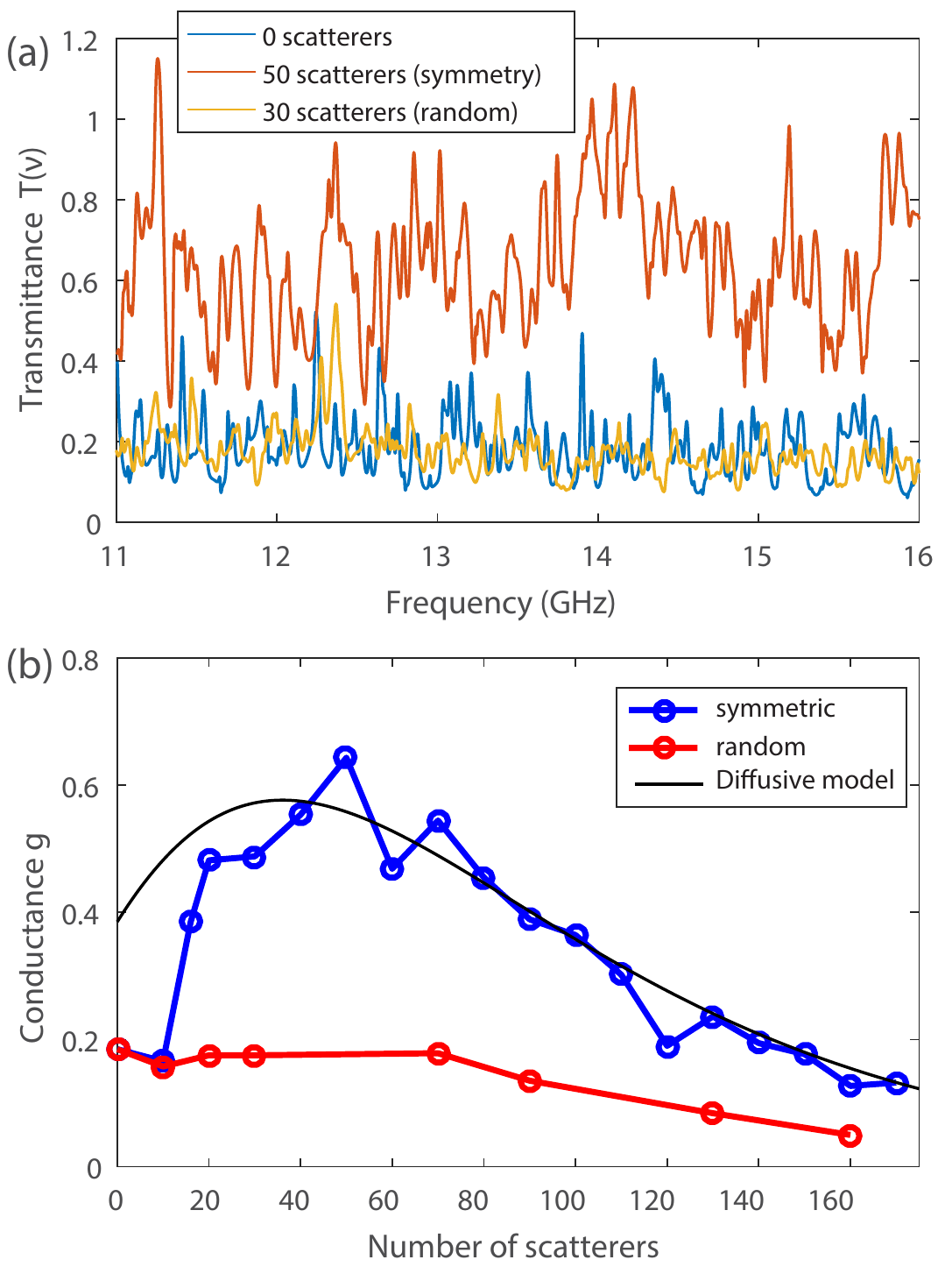}
\caption{\label{fig_scatterers} (a) Spectra of the transmittance $T(\nu)$ for a $7.8$mm-high barrier, in absence of scatterers (blue line), with $N_\textup{s} = 50$ aluminum cylinders placed randomly but with left-right symmetry (orange line) and 30 cylinders placed fully randomly. (b) Variation of the conductance $g(N_\textup{s})$ with respect to the number of randomly located scatterers $N_\textup{s}$ in a symmetric (blue line) or fully random (orange line) arrangement.}
\end{figure}

After testing the impact of the barrier alone, we show that the conductance can be further enhanced by introducing a diffusive medium with a symmetric arrangement. We place a collection of $N_\textup{s}$ aluminum cylinders on both sides of the $7.8$~mm high barrier (see Fig.~\ref{fig:fig1}(b)). The transmittance $T(\nu)$ is shown in Fig.~\ref{fig_scatterers}(a) for $N_\textup{s} = 0$, $N_\textup{s} = 30$ for two independent random configurations on both sides of the barrier, and $N_\textup{s} =50$ for a random arrangement with a mirror symmetry. For a non-symmetric configuration, the system is slightly more opaque than for a barrier alone as the disorder strength has increased. However, a clear broadband enhancement is observed for symmetrically placed metallic scatterers. $g(N_\textup{s})$ indeed increase with $N_\textup{s}$, reaches a maximum value of $g=0.64$ for $N_\textup{s}=50$ and then decreases with $N_s$ as it is expected to vanish in the strong disorder limit $N_\textup{s} \rightarrow \infty$ (see Fig.\ref{fig_scatterers}(b)). In contrast, in absence of mirror symmetry $g(N_\textup{s})$ only decreases with $N_\textup{s}$ as a result of the combined effect of disorder and the barrier.

The experimental results are now compared to a diffusive model proposed for disordered waveguides with open boundary conditions at left and right surfaces \cite{Cheron2020_scirep}. We do not expect that this model can accurately describe our experimental results for small $N_\textup{s}$, as it does not account for the strong internal reflections at both interfaces. For large $N_\textup{s}$, however, the main process is the diffusion within the scattering disorder. The key parameters are the barrier strength $\varsigma_\textup{B}=g_0/N$, with $g_0$ the conductance in absence of disorder, the normalized sample length $s = L/\ell$, with $\ell$ the mean free path, and $s_\textup{a} = \sqrt{\ell_\textup{a}/(2\ell)}$, with $\ell_\textup{a}$ the ballistic absorption length. A scaling model including losses gives the following expression for the theoretical conductance $g_{\textup{theo}}(s,s_\textup{a},\varsigma_\textup{B})$:
\begin{equation}
    \frac{1}{g_\textup{theo}(s,s_\textup{a},\varsigma_\textup{B})}= \frac{\alpha(s,s_\textup{a},\varsigma_\textup{B})}{g_\textup{B}(s,s_\textup{a})} + \frac{1}{g_\textup{E}(s,\varsigma_\textup{B})}.
    \label{eq:conductance_scatterers}
\end{equation}
Here, $g_\textup{B}(s,s_\textup{a}) = 2 N s_\textup{a}^{-1} e^{-s/s_\textup{a}}$ corresponds to the absorbing diffusive conductance in the absence of the barrier and $\alpha(s,s_\textup{a},\varsigma_\textup{B}) = 1+(2 s_\textup{a}^2 \varsigma_\textup{B})^{-1}$. $g_\textup{E}(s,\varsigma_\textup{B})$ reflects the enhancement of the conductance due to mirror symmetry, $g_\textup{E}(s,\varsigma_\textup{B}) = [1 + s \varsigma_\textup{c} / (1-\varsigma_\textup{c})] N \varsigma_\textup{B} / (1-\varsigma_\textup{B})$, where $\varsigma_\textup{c} \simeq 0.4$ is a parameter that has been found from a fit of $g$ with $s$ in numerical simulations.

In empty chaotic cavities the conductance is equal to $N/2$. We therefore estimate that the conductance associated to the barrier in absence of internal reflections is $g_0 \sim 2 g(N_\textup{s}=0)=0.4$ giving $\varsigma_\textup{B} = 0.05$. We then use that 1) $\ell$ scales linearly with $N_\textup{s}$, $s=\kappa N_\textup{s}$, and 2) the optimal conductance is found for $N_\textup{s} = N_\textup{opt}$ so that $\kappa = s_\textup{opt}/N_\textup{opt}$ with, in the absence of absorption, $s_\textup{opt} = \sqrt{(\varsigma_\textup{B}^{-1}-1)(\varsigma_\textup{c}^{-1}-1)} -(\varsigma_\text{c}^{-1}-1)$. Our experimental results shown in Fig.~\ref{fig_scatterers}(b) present a maximum of $g$ for approximatively 50 scatterers but strong fluctuations on the data are observed around this optimal value. Moreover, $N_\textup{opt}$ should be its value in the absence of absorption and is therefore larger than $N_\textup{s} = 50$. We estimate here that $N_\textup{opt} = 80$ is reasonable value that provides a good agreement with the data. The last parameter $s_\textup{a} = 3.6$ is finally obtained from the best fit of the tail of $g(N_\textup{s})$ as $N_\textup{s} > 100$. The theoretical curve is in a good agreement with measurements in Fig.~\ref{fig_scatterers}(b) for $N_\textup{s}>30$. We stress that the enhancement of the conductance is even stronger in our system relative to open random waveguides as the transmission for a barrier alone is reduced by a factor $\sim 1/2$ due to strong internal reflections at the metallic boundaries of the cavity. \\

In conclusion, we have provided a clear experimental observation of the impact of interference effects due to mirror symmetry on the transmission through disordered multichannel cavities. For a barrier placed within a cavity with strong internal reflections at its interfaces, constructive interference yields a broadband enhancement of the fraction of open transmission eigenchannels and consequently of the conductance. A further increase of the conductance has been obtained by symmetrically placing scatterers around the barrier, with experimental results in good agreement with a theoretical model including losses. The sensitivity of these systems may also open up new perspectives to detect defects within complex structures.\\

\section{Supplementary Material}
See supplemental material for numerical simulations of the distribution of transmission eigenvalues in random samples with absorption.

\section{Acknowledgments}
This publication was supported by the European Union through the European Regional Development Fund (ERDF), by the French region of Brittany and Rennes M{\'e}tropole through the CPER Project SOPHIE/STIC \& Ondes. M. D. acknowledges the Institut Universitaire de France. C. F. acknowledges funding from the French "Minist{\`e}re de
la D{\'e}fense, Direction G{\'e}n{\'e}rale de l'Armement''.  E. C. acknowledges funding by the project HYPERMETA funded under the program Etoiles Montantes of the Region Pays de la Loire.

The data that support the findings of this study are available from the corresponding author upon reasonable request.

\bibliographystyle{apsrev4-1}

\providecommand{\noopsort}[1]{}\providecommand{\singleletter}[1]{#1}%

\clearpage

\renewcommand{\thefigure}{S\arabic{figure}}
\renewcommand{\theequation}{S\arabic{equation}}
\setcounter{equation}{0}
\setcounter{figure}{0}
\setcounter{section}{0}

\date{\today}

\title{Supplemental Material for 'Experimental evidence of enhanced broadband transmission in disordered systems with mirror symmetry''}
\begin{titlepage}
\maketitle
\end{titlepage}

\section{Supplemental Material for 'Experimental evidence of enhanced broadband transmission in disordered systems with mirror symmetry'}

In this Supplementary Material, we explore the impact of losses on the distribution of transmission eigenvalues $P(\tau)$ in simulations of waves propagating through lossy random media. Our experimental results presented in Fig.~2b of the main text indeed show that the peak corresponding to open channels is shifted from $\tau_m=1$ expected theoretically in absence of absorption \cite{Baranger1994,Jalabert1994,Goetschy2013} to $\tau_m = 0.9$ and we attribute this effect to the impact of inevitable losses. We consider a two-dimensional waveguide of width $W$ and length $L =2W$ with reflective transverse boundaries. The boundary conditions are opened at the right and left surfaces. 
\begin{figure}
\includegraphics[width=8.5cm]{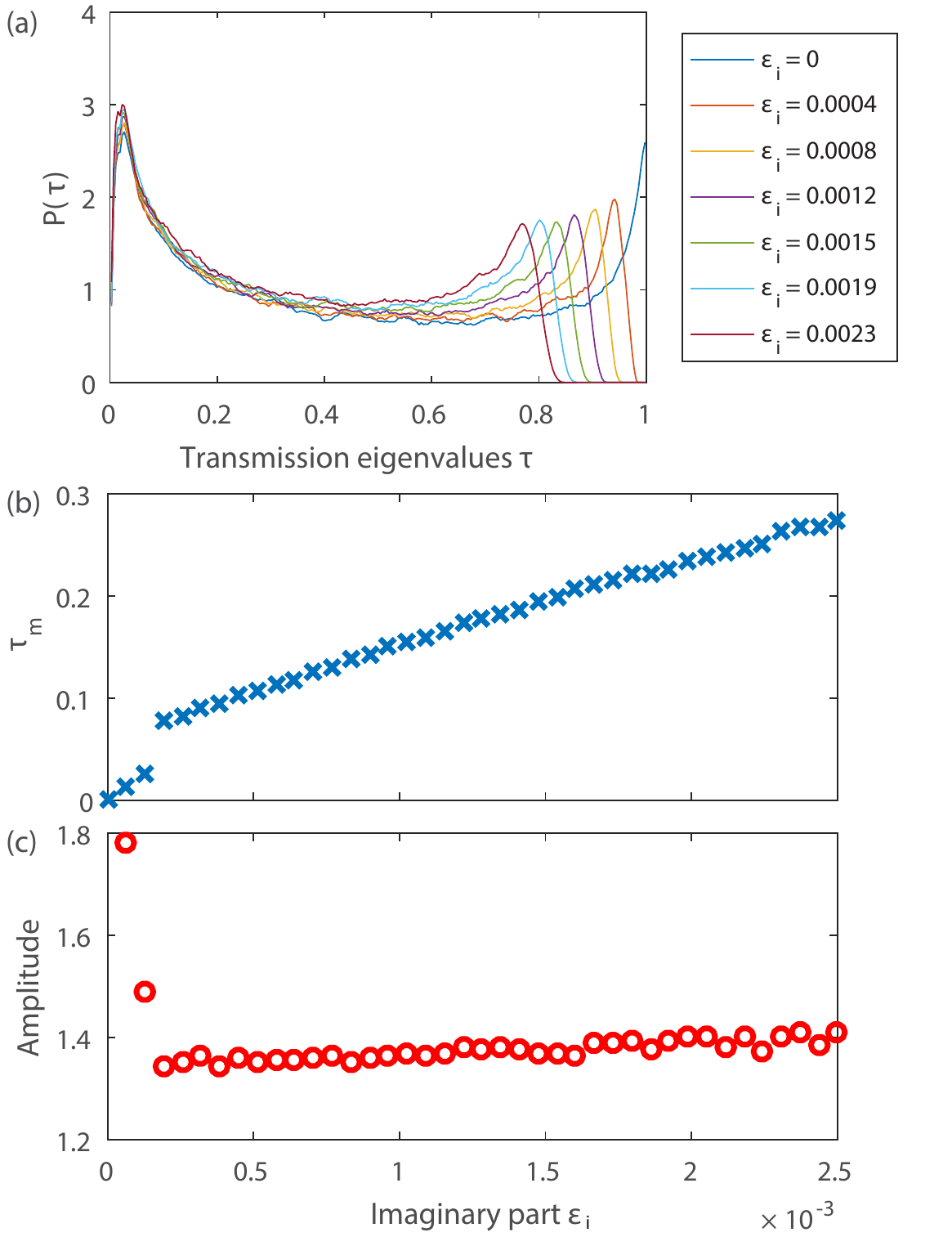}
\caption{\label{fig:fig1} (a) Distribution of transmission eigenvalues of a random waveguide with $N=8$ channels for different values of the imaginary part of the dielectric permittivity. The second characteristic peak shifts from $\tau_m = 1$ to smaller values as $\epsilon_i$ increases. (b,c) Variations of $\tau_m$ (b) and amplitude (c) of the second characteristic peak of $P(\tau)$ with  $\epsilon_i$.}
\end{figure}
The wavelength $\lambda$ is chosen so that the empty waveguide supports $N=8$ channels ($W = 4\lambda$). The Green's functions between points at the input and output interfaces of a random waveguide are first obtained by solving the two-dimensional wave equation $\nabla^2 \psi(x,y) +k_0^2 \epsilon(x,y) \psi(x,y) = 0 $ using the recursive Green’s function method \cite{Baranger1994}. The dielectric permittivity $\epsilon(x,y)$ is a random asymmetrical function drawn from a rectangular distribution. We add a constant imaginary part $\epsilon_i$ representing absorption to $\epsilon(x,y)$. The elements of the transmission matrix $t_{ba}$ between incoming modes $a$ and outgoing modes $b$ are then calculated by projecting the Green’s functions onto the modes of the empty waveguide. The transmission eigenvalues $\tau_n$ are found from a diagonalization of $t^\dagger t$, $t^\dagger t = \Sigma_{n=1}^N v_n \tau_n v_n^\dagger$ with $v_n$ being the corresponding eigenvectors.

The distribution of transmission eigenvalues $P(\tau)$ is found from an ensemble of 4000 random waveguides. The variance of $\epsilon(x,y)$ is chosen so that the average transmission $\langle T \rangle = \langle \tau \rangle = 0.44$ in the absence of absorption. In this case, $P(\tau)$ is bimodal with two peaks at $\tau = 0$ and $\tau = 1$ corresponding to closed and open channels, respectively. However, the second characteristic peak shifts towards smaller transmission $\tau_m$ as absorption within the samples increases (see Fig.~\ref{fig:fig1}(a)), in agreement with our experimental result. The transmission $\tau_m$ corresponding to this peak increases monotonically with $\epsilon_i$ as seen in Fig.~\ref{fig:fig1}(b) with an amplitude first rapidly decreasing with absorption strength and then saturating (see Fig.~\ref{fig:fig1}(c)).

Even though the geometry of the cavity considered experimentally is different from the case of random waveguides since the empty cavity features strong internal reflections at left and right interfaces instead of internal disorder, these simulations demonstrate that small internal losses do not suppress the second peak on the distribution of transmission eigenvalues but rather leads to a shift of this peak.  


\bibliographystyle{apsrev4-1}
\providecommand{\noopsort}[1]{}\providecommand{\singleletter}[1]{#1}%

\end{document}